\documentclass[preprint2]{aastex}		% Two-column preprint style

\def\sun{_\odot}
\def\kms{\,km\,s$^{-1}$}

\def\Dwa{$\,$\uppercase\expandafter{\romannumeral5}$\,$}

\def\sles{\lower2pt\hbox{$\buildrel {\scriptstyle <}
   \over {\scriptstyle\sim}$}}

\def\sgreat{\lower2pt\hbox{$\buildrel {\scriptstyle >}
   \over {\scriptstyle\sim}$}}
\def\sharpnull#1{}

\newcommand\hii{H{\sc ii}}

\def\ls{\mathrel{\mathchoice {\vcenter{\offinterlineskip\halign{\hfil
$\displaystyle##$\hfil\cr<\cr\sim\cr}}}
{\vcenter{\offinterlineskip\halign{\hfil$\textstyle##$\hfil\cr
<\cr\sim\cr}}}
{\vcenter{\offinterlineskip\halign{\hfil$\scriptstyle##$\hfil\cr
<\cr\sim\cr}}}
{\vcenter{\offinterlineskip\halign{\hfil$\scriptscriptstyle##$\hfil\cr
<\cr\sim\cr}}}}}
\begin{document}

\shorttitle{{\sl Swift} Observations of SN~2005cs}
\shortauthors{Brown et~al.}

%%%%%%%%%%%%%%%%%%%%%%%%%%%%%%%%%%%%%%%%%%%%%%%%%%%%%%%%%%%%%%%%%%%%%%%%%%%%%%
%%%%%%%%%% Title & Author %%%%%%%%%%%%%%%%%%%%%%%%%%%%%%%%%%%%%%%%%%%%%%%%%%%%

\title{Early Ultraviolet, Optical and X-Ray Observations \\of the Type IIP 
SN~2005\lowercase{cs} in M51 with {\sl Swift}}

\author{Peter J.~Brown\altaffilmark{1,2},
	Luc Dessart\altaffilmark{3},
        Stephen~T.~Holland\altaffilmark{4,5},
	Stefan Immler\altaffilmark{4,5},
	Wayne Landsman\altaffilmark{4},
	St\'{e}phane Blondin\altaffilmark{6},
	Alexander~J.~Blustin\altaffilmark{7},
	Alice Breeveld\altaffilmark{7},
	Gulab~C.~Dewangan\altaffilmark{8},
	Neil Gehrels\altaffilmark{9},
	Robert B.~Hutchins\altaffilmark{6},
	Robert P.~Kirshner\altaffilmark{6},
	Keith O.~Mason\altaffilmark{7},
	Paolo A.~Mazzali\altaffilmark{10,11},
	Peter Milne\altaffilmark{3},
	Maryam Modjaz\altaffilmark{6}, \&
        Peter W.~A.~Roming\altaffilmark{1}
	}

\altaffiltext{1}{Pennsylvania State University,
                 Department of Astronomy \& Astrophysics,
                 University Park, PA 16802, USA}
                 
\altaffiltext{2}{pbrown@astro.psu.edu}                

\altaffiltext{3}{Department of Astronomy and Steward Observatory, 
		University of Arizona, 
		Tucson, AZ 85721, USA}

\altaffiltext{4}{Astrophysics Science Division,
		X-Ray Astrophysics Branch, Code 662,
                 NASA Goddard Space Flight Center,
                 Greenbelt, MD 20771, USA}

\altaffiltext{5}{Universities Space Research Association,
		10211 Wincopin Circle, Columbia MD 21044, USA}

\altaffiltext{6}{Harvard-Smithsonian Center for Astrophysics,
		Cambridge, MA 01238, USA}

\altaffiltext{7}{Mullard Space Science Laboratory,
		 Department of Space and Climate Physics,
		 University College London, Holmbury St Mary,
		 Dorking, Surrey, RH5 6NT,
		 UK}

\altaffiltext{8}{Department of Physics,
		Carnegie Mellon University,
		5000 Forbes Avenue,
		Pittsburgh, PA 15213, USA}

\altaffiltext{9}{Astrophysics Science Division,
		Astroparticle Physics Laboratory, Code 661,
                 NASA Goddard Space Flight Center,
                 Greenbelt, MD 20771, USA}

\altaffiltext{10}{INAF, Osservatorio Astrnomico di Trieste, 
		via G.B. Tiepolo 11, 34131 Trieste, Italy}

\altaffiltext{11}{Max-Planck-Institut fur Astrophysik, 
		Karl-Schwarzschild Strasse 1, 85741, 
		Garching, Germany}

%%%%%%%%%% Abstract %%%%%%%%%%%%%%%%%%%%%%%%%%%%%%%%%%%%%%%%%%%%%%%%%%%%%%%%%

\begin{abstract}

We report early photospheric-phase observations of the Type IIP 
Supernova (SN) 2005cs obtained by {\sl Swift}'s Ultraviolet-Optical  
and X-Ray Telescopes. Observations started within two days of discovery 
and continued on a regular basis for three weeks.
During this time the V-band magnitude remained essentially constant, 
while the UV was initially bright but steadily faded until below the brightness 
of an underlying UV-bright H{\sc ii} region.  This UV decay is similar 
to SNe II observed by the International Ultraviolet Explorer.  
UV grism spectra show the P-Cygni absorption of Mg{\sc ii} 2798\AA, 
indicating a photospheric origin of the UV flux.
Based on non-LTE model atmosphere calculations with the CMFGEN code, we associate 
the rapid evolution of the UV flux with the cooling of the 
ejecta, the peak of the spectral energy distribution (SED) 
shifting from $\sim$700\AA\ on June 30th
to $\sim$1200\AA\ on July 5th. Furthermore, the corresponding recombination of 
the ejecta, e.g., the transition from Fe{\sc iii} to Fe{\sc ii}, 
induces a considerable strengthening of metal line-blanketing at and 
above the photosphere, blocking more and more effectively this fading UV flux. 
SN~2005cs was not detected in the X-rays, and the upper limit to the X-ray luminosity yields a 
limit to the mass loss rate of the progenitor of 
$\dot{M} \ls 1 \times 10^{-5}~M_{\odot}~{\rm yr}^{-1}~(v_{\rm w}/10~{\rm km~s}^{-1})$.
Overall, {\sl Swift} represents a unique opportunity to capture the early and fast 
evolution of Type II SNe in the UV, providing additional constraints on the reddening, 
the SED shortward of 4000\AA, and the ionization state and temperature of the 
photon-decoupling regions. 
\end{abstract}

%%%%%%%%%% Keywords %%%%%%%%%%%%%%%%%%%%%%%%%%%%%%%%%%%%%%%%%%%%%%%%%%%%%%%%%


%%%%%%%%%%%%%%%%%%%%%%%%%%%%%%%%%%%%%%%%%%%%%%%%%%%%%%%%%%%%%%%%%%%%%%%%%%%%%

\section{Introduction\label{SECTION:intro}}

     Supernova (SN) 2005cs was discovered in the Whirlpool Galaxy (M51, NGC 5194) by
W. \citet{K05} on 2005 June 28.9 (all dates UT).  \citet{Pastorello_etal_2006} estimate the explosion 
occurred June 27.5 (JD 2453549) based on previous non-detections.  
\citet{MKC05} classified it from a spectrum as a young SN II on June
30.23, and the observations with the {\sl Swift} spacecraft \citep{G05} reported
here began June 30.9.  The photometric and spectroscopic evolution of SN~2005cs 
is consistent with that of a subluminous SN II-Plateau (IIP), with a plateau phase 
lasting $\sim110$ days and low ejecta velocities (\citealp{Tsv_2006};
\citealp{Pastorello_etal_2006}).

A young SN discovered in a nearby galaxy as well studied as M51 offers a unique
opportunity for detailed investigations.  The host galaxy has previously been 
observed at various wavelengths by many different instruments (see e.g. 
\citealt*{Cal05},  \citealt*{Dew05}).  Examination of pre-explosion 
images from the Hubble Space Telescope (HST) has identified a $7-9 M_{\sun} $ red 
supergiant as a plausible progenitor candidate (\citealt*{Li06}; \citealt*{MSD05}).
The position measured using high resolution images from the
Canada-France-Hawaii Telescope and HST is 
R.A. $=13^\mathrm{h}29^\mathrm{m}52\fs764$ and Dec. $=+47\arcdeg10\arcmin36\arcsec09$ 
(J2000; \citealt*{Li06}).

The quick identification and classification of this SN event allowed 
for early observations by instruments on the ground and in space while the 
SN was still in its early photospheric phase.   
This is especially critical in the UV 
because of the rapid decline in luminosity at short wavelengths as the SN photosphere cools.  Because they require 
space-based observatories, UV data on SNe are limited, especially at early times.  
While IUE observed several SNe II, none were of the ''plateau'' (IIP) subtype despite it being the
most common type of SNe II.  
HST observed the SN IIP 1999em at two epochs, starting about two weeks after the 
explosion \citep{Baron_etal_2000}.  
The observations presented in this paper fill an important gap 
by showing the UV evolution of SN~2005cs during the first few weeks after its explosion.

%%%%%%%%%%%%%%%%%%%%%%%%%%%%%%%%%%%%%%%%%%%%%%%%%%%%%%%%%%%%%%%%%%%%%%%%%%%%%%

\section{Observations and Reductions\label{SECTION:obs}}

\subsection{{\sl Swift} Observations\label{SECTION:phot_01}}

\begin{figure}
\epsscale{1.0}
\plotone{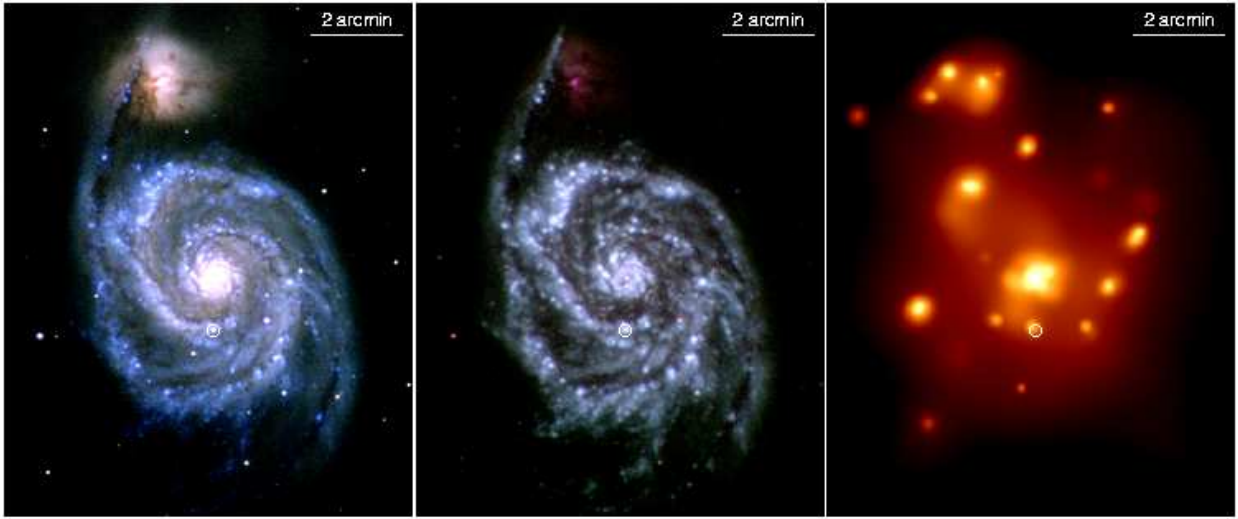}
\caption{  {\sl Swift} UVOT optical (left), UV (middle) and XRT X-ray (right) images 
of SN~2005cs and its host galaxy M51. The position of SN~2005cs is indicated 
by a white circle of $8''$ radius and the spatial scale, identical for each image, 
is indicated in the top corner of each panel.
The optical image was contsructed 
from summed images from the UVOT $V$ (1,815~s exposure time; red), $B$ (1,232~s; green), and 
$U$ (1,065~s; blue) filters, the UV image from the UVOT UVW1 (3,038~s 
red), UVM2 (7,189~s; green) and UVW2 (4,703~s; blue) filters  
and slightly smoothed with a Gaussian filter of 1.5 
pixel (FWHM).  The (0.2--10~keV) XRT X-ray image was constructed from the merged 
33~ks XRT data and adaptively smoothed using the CIAO command csmooth to 
achieve a S/N in the range 2.5 to 4.}\label{fig_UVM51}
\end{figure}

     Regular observations of SN~2005cs were made with the {\sl Swift}
spacecraft \citep{G05} between 2005 June 30th and July 23rd with a follow-up  
observation on September 14, utilizing both the X-Ray Telescope (XRT; \citealp*{Bur05})
and the Ultraviolet/Optical Telescope (UVOT; \citealp{R05}).  Images from UVOT and XRT 
of SN~2005cs and its host galaxy M51 are displayed in Fig.~\ref{fig_UVM51}.  The XRT operates 
in the 0.2--10 keV range with a $18''$ half-power diameter at 1.5 keV.  
%The UVOT is a 30 cm telescope with a photon counting detector.  
Pertinent calibration details for UVOT\footnote{see http://swift.gsfc.nasa.gov/docs/heasarc/caldb/swift/docs/uvot/ 
for updated documentation}, 
including the filter bandpasses and photometric zeropoints, are given in Table 1. 
% The first observations were made by uploading the target to the {\sl Swift} 
%spacecraft as a Target of Opportunity.

%%%%%%%%%%%%%%%%%%%%%%%%%%%%%%%%%%%%%%%%%%%%%%%%%%%%%%%%%%%%%%%%%%%%%%%%%%%%%%%%%

\subsection{UVOT Photometry\label{SECTION:phot_02}}

      SN~2005cs is located in a spiral arm of M51,
which causes the background to be variable over small spatial scales.  
There is an underlying \hii~region visible in pre-explosion GALEX UV images \citep{Bia04}.
This means that high precision
photometry will not be possible until SN~2005cs has faded and the
underlying light can be subtracted.  It is anticipated that {\sl Swift} 
UVOT will reobserve the field of SN~2005cs in 2007 so
that late-time template images of the host galaxy can be obtained.  
At the early times in which we are most interested here, the SN was much brighter 
and contribution from the \hii~region or other contamination is small.

      We performed aperture photometry using a circular aperture with a
radius of $2\arcsec$ ($UBV$) or $4\arcsec$ (UV filters) centered on the SN.  
A background sky region was selected by
eye which contained approximately the same structure and light as the
region containing the SN.  Photometry was done using several
background regions selected in this way and all were found to return
similar photometry.  Aperture corrections to the standard UVOT photometry apertures
($6\arcsec$ for $UBV$ and $12\arcsec$ for UVW1, UVM2, and UVW2) were calculated
frame by frame in $UBV$.  For the UV filters, a single aperture correction
for each filter was computed from a summed image due to the lack of UV-bright,
isolated stars in the field of view.  The larger source aperture was chosen to 
minimize errors due to small orbital variations in the PSF.

\begin{figure}
\epsscale{.6}
\plotone{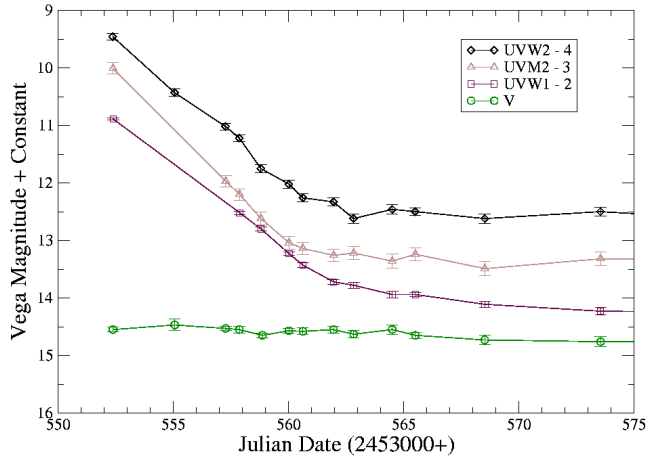}
\caption{Light curves of SN~2005cs obtained by UVOT in the three UV filters and 
the V band.  For clarity, the UV curves have been shifted by a constant offset given in the legend.  
The steep decay in the UV levels off as the SN fades below the 
brightness of the underlying \hii~region.}
\label{fig_lc}
\end{figure}

The count rates were corrected for coincidence loss using the coincidence loss equation in the 
{\sl Swift} UVOT Calibration Database
(CalDB)\footnote{available from http://swift.gsfc.nasa.gov/docs/heasarc/caldb/swift/}.  
Corrected count rates were 
transformed to Vega magnitudes using the appropriate photometric zero points in the CalDB.
The $U$ and $B$ curves begin near or above the saturation level of UVOT's photon 
counting detector and will not be considered here.  The
photometry in the remaining four UVOT filters, 3 UV filters and the V band, is presented 
in Table~2 and displayed in Fig.~\ref{fig_lc}.  These values have not been 
corrected for extinction.  The errors 
(given in Table 2 or displayed in Fig.~\ref{fig_lc}) are the $1\sigma$ statistical errors only and do not include 
the systematic errors in the photometric zeropoint calibration (given in Table 1).

%%%%%%%%%%%%%%%%%%%%%%%%%%%%%%%%%%%%%%%%%%%%%%%%%%%%%%%%%%%%%%%%%%%%%%%%%%%%%%

\subsection{UVOT Grism Data}

Swift UVOT also observed SN~2005cs several times using its 
spectroscopic grisms.  These observations are listed in 
Table 3. The wavelength scale 
could be shifted by up to 30 \AA~due to the difficulty of 
fixing the scale on the saturated zeroth order spectrum especially at early times when 
the UV flux was greatest.  There is also a large uncertainty in the 
background subtraction due to the bright underlying galaxy.   
  
\begin{figure}
\epsscale{.6}
\plotone{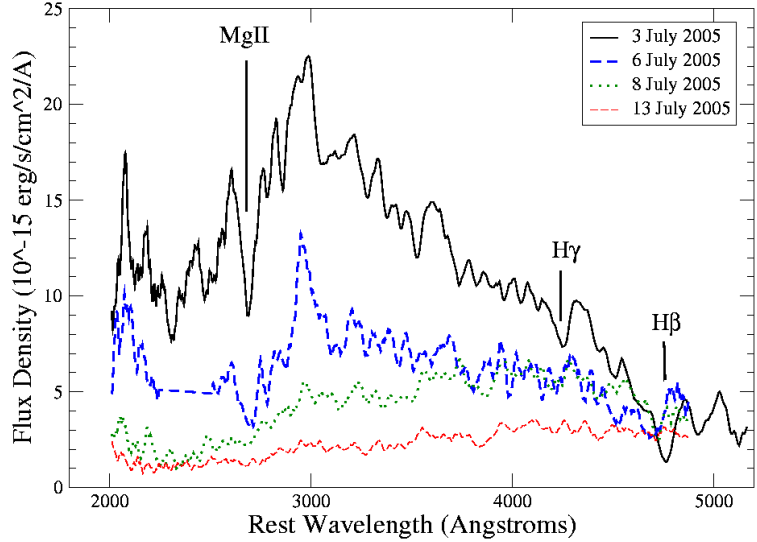}
\caption{Grism spectra of SN~2005cs obtained by UVOT, smoothed and scaled to 
contemporaneous UVOT photometry.  The first spectrum is a composite of UV and 
V grism spectra (spliced at 2900\AA) and the others from the UV grism.  
We highlight a few features, which we associate with the absorption 
part of P-Cygni profiles for Mg{\sc ii}\,2798\AA, H$\beta$ and H$\gamma$:
these Doppler-broadened absorptions are consistent with other optical features,
whose modeling supports a formation within the fast-expanding SN ejecta, rather than
in the CSM (See \S~3.3 for discussion). }\label{fig_grism}
\end{figure}

The extracted spectra were smoothed using a running average of 10 points 
($\sim$20\AA) and scaled to match contemporaneous 
UVOT photometry (using the flux density conversion factors of Table 1).  
The resulting spectra from four epochs are 
displayed in Fig.~\ref{fig_grism}.   The first epoch spectrum is a 
composite of the UV grism and V grism spectra, 
spliced near 2900\AA, in order to avoid order overlap in the UV grism. 
The UV grism observation from 6 July contains the zeroth order 
of another field star near 2450 \AA, and the affected region 
has been removed from the spectrum.

%%%%%%%%%%%%%%%%%%%%%%%%%%%%%%%%%%%%%%%%%%%%%%%%%%%%%%%%%%%%%%%%%%%%%%%%%%%%%%

%%%%%%%%%%%%%%%%%%%%%%%%%%%%%%%%%%%%%%%%%%%%%%%%%%%%%%%%%%%%%%%%%%%%%%%%%%%%%%
\subsection{XRT Data}\label{xrt}

All XRT data, from June 30th to September 14, were merged into a single 33.0~ks observation to search
for X-ray emission from SN~2005cs. However, the bright
(integrated flux between 0.2 and 10 keV, $f_{0.2-10}=[3.3\pm0.7] \times 10^{-14}~{\rm ergs~cm}^{-2}~{\rm s}^{-1}$)
and nearby ($8''$ offset) X-ray flash source detected in sequence number
00030083011 (Immler, Kong \& Lewin 2005) severely affects an estimate of
the amount of emission from SN~2005cs because this source is within the
XRT source extraction aperture ($18''$ half-power diameter at 1.5 keV).
We therefore excluded all observations obtained on the day the X-ray
flash was observed (2005-07-06, sequence numbers 00030083010/11/12/13) to
minimize contamination, leaving 26.8~ks of exposure time in the final 
merged image used below.

The 26.8~ks image shows no X-ray
source at the position of the X-ray flash, and no source at the
position of SN~2005cs. We therefore extracted on-source counts from a
circular extraction region with a radius of 10~pixel ($23\farcs6$) and
subtracted the background from an annulus centered on the position of
SN~2005cs to account for diffuse emission from the galaxy.
A 3-sigma upper limit to the 0.2--10~keV count rate of
$1.7\times10^{-3}~{\rm cts~s}^{-1}$ is obtained, which corresponds to
an unabsorbed X-ray flux of
$f_{0.2-10}\ls 9.0 \times 10^{-14}~{\rm ergs~cm}^{-2}~{\rm s}^{-1}$.

%%%%%%%%%%%%%%%%%%%%%%%%%%%%%%%%%%%%%%%%
%%%%%%%%%%%%%%%%%%%%%%%%%%%%%%%%%%%%%%%%%%%%%%%%%%%%%%%%%%%%%%%%%%%%%%%%%%%%%%

\subsection{{\sl XMM-Newton} EPIC Data}
\label{epic}

SN~2005cs was observed with XMM-Newton on 2005-07-01 (OBS-ID 0212480801)
as a Target of Opportunity (PI Immler) to search for prompt X-ray emission
from the SN. Data processing and analysis were performed with
SAS~6.5.0\footnote{http://xmm.vilspa.esa.es/external/xmm\_sw\_cal/sas.shtml}
and the latest calibration constituents. After screening of the EPIC PN and
MOS data for periods with a high background, cleaned exposure times of
28.1~ks for the PN and 33.8~ks for each of the two MOS instruments were obtained.

Inspection of the EPIC images showed that the X-ray ``flash'' serendipitously
observed by {\sl Swift} XRT during some of the observations (Immler, Kong \& Lewin 2005) 
is also present in the data obtained quasi-simultaneously with the XMM-Newton EPIC.
Since the offset of this bright X-ray source from the position of SN~2005cs
($8''$) is smaller than the point-spread-function of the XRT+EPIC ($15''$
half energy width), no reliable upper limits to the X-ray flux of SN~2005cs
can be established from the XMM-Newton X-ray data. The XMM-Newton EPIC
data will therefore not be further used and we will rely on the {\sl Swift}
XRT data to study the X-ray emission from SN~2005cs.

%%%%%%%%%%%%%%%%%%%%%%%%%%%%%%%%%%%%%%%%%%%%%%%%%%%%%%%%%%%%%%%%%%%%%%%%%%%%%%
\section{Discussion\label{SECTION:discussion}}
%%%%%%%%%%%%%%%%%%%%%%%%%%%%%%%%%%%%%%%%%%%%%%%%%%%%%%%%%%%%%%%%%%%%%%%%%%%%%%
\subsection{Ultraviolet Light Curves\label{SECTION:lightcurves}}

SN~2005cs was initially extremely bright in the UV, outshining the nucleus of M51 
by over 2 magnitudes.  The blue UV-optical colors 
of SN~2005cs were used to photometrically type SNe~2006at \citep{BI06} and 
2006bc \citep{IB06} as young SNe II.  The UV faded quickly 
until falling below the brightness of an underlying \hii~region about 10 days after the explosion.
While the decay slope generally steepens at shorter wavelengths (see \citealp{Li06}, 
\citealp{Tsv_2006}, or \citealp{Pastorello_etal_2006} for the optical curves), 
the decline in UVM2 is actually steeper (0.38 mag/day) than the filters on 
either side, UVW1 (0.31 mag/day) UVW2 (0.34 mag/day).  This is likely caused 
by the strong Fe{\sc iii} and the 
strengthening Fe{\sc ii} lines concentrated within the UVM2 bandpass.

\begin{figure}
%\includegraphics[scale=1]{f4_color.eps}
%\begin{figure}
\epsscale{.6}
\plotone{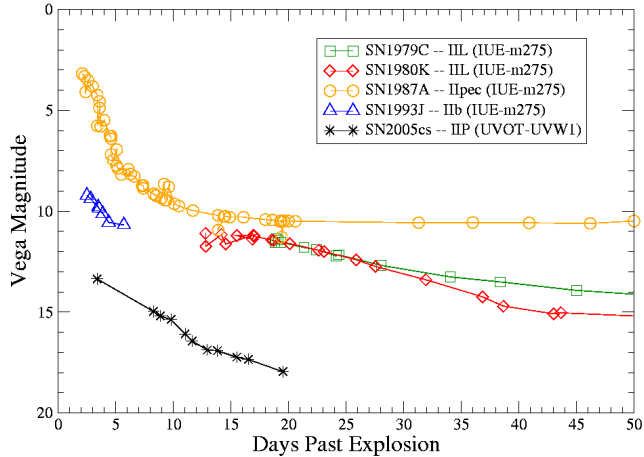}
\caption{Comparison of SNe II observed by IUE and {\sl Swift}-UVOT.  An early UV decay 
is seen in all these SNe despite the differences in subclass or optical behavior.}\label{fig_lc_comp}
\end{figure}

For a comparison with other SNe II in a nearby UV wavelength regime, 
we created a new UVW1 light curve for SN~2005cs by assuming that 
the magnitude in the last epoch ($\sim80$ days after explosion) corresponds to 
the underlying background and subtracting that 
flux from the early UVW1 data.  This might overestimate the background caused by 
a residual contribution from the SN, but this is not expected to be significant  
because of the continued fading seen in more recent UV observations of SNe II 
such as SN~2006bp \citep{Immler_etal_2006}.  
This new UVW1 curve is compared to other UV observations of SNe II in Fig.~\ref{fig_lc_comp}.  

Several SNe II were observed by IUE, including several with multiple epochs
(more than four observations: SNe 1979C, 1980K, 1987A \& 1993J).  
For comparison we use the $m_{275}$ magnitudes given by \citet{CTF} 
which are calculated by convolving the HST F275W 
filter (central wavelength 2770\AA~and width 594\AA; \citealp{Nota_etal_1996}) 
with the IUE spectra and calibrated so Vega would have $m_{275}=0$.  Due to 
the quantity of data, only an early selection from SN~1987A was used. 
Any magnitudes from spectra marked as saturated were also left out, 
as were the later spectra of SN~1993J which were contaminated by sunlight.  
We left the IUE magnitudes 
unshifted, as the difference in apparent magnitude between the brighter SNe observed by 
IUE and SN~2005cs is enough to separate their respective curves.

To make sure the comparison of the UV light curve shapes with the UVW1 and 
$m_{275}$ filters is valid, we performed spectrophotometry 
on the IUE spectra of SN~1987A \citep{Pun_etal_1995} using the IRAF SBANDS command.  
A selection of spectra from the Long Wave Prime (LWP) camera 
(1975 \AA~--3300\AA; \citealp{Pun_etal_1995}) 
were convolved with the UVW1 effective area curves 
to simulate how SN~1987A would have appeared in the UVW1 filter.  The light 
curves of the UVW1 spectrophotometry and $m_{275}$ were very similar.

%%%%%%%%%%%%%%%%%%%%%%%%%%%%%%%%%%%%%%%%%%%%%%%%%%%%%%%%%%%%%%%%%%%%%%%%%%%%%%
While the fading behavior in the UV is common among these SNe, it is interesting to 
note how varied the early optical ($V$ band) behavior is for these SNe during the 
same time period.  SN~1979C 
(\citealp{Panagia_etal_1980}, \citealp{BCR_1982}) 
and SN~1980K \citep{But82}, both II-Linear SNe (IIL), were discovered after 
maximum light and were thus fading in the optical.  
SN~1993J \citep{Ric94}, a IIb, was caught soon after explosion and showed the  
brightness declining, reaching a local minimum about 8 days
before rising again as the photosphere receded to layers influenced by non-thermal
excitation due to the radioactive decay of unstable isotopes of Nickel and Cobalt.  
SN~1987A \citep{Pun_etal_1995}, a peculiar type II,  
was mostly rising during this period.  
As shown in Fig.~\ref{fig_lc}, SN~2005cs had a nearly 
constant V magnitude during this period, the distinguishing characteristic 
of a SN II-Plateau. In all cases, the UV and optical light curves must
have begun with a rapid rise at the time of shock breakout, which is not usually 
observed.  But while the optical curves vary enough to separate SNe II into 
subclasses based on their lightcurve shape \citep{BCR_1979}, the differences 
from subclass or emission mechanism are not as clear in the UV.

%The addition of UV observations of SN~2005cs to this group 
%is important, as IIP SNe are the most common type of SNe II, 
%but their temporal behavior in the UV was previously not well observed.

%%%%%%%%%%%%%%%%%%%%%%%%%%%%%%%%%%%%%%%%%%%%%%%%%%%%%%%%%%%%%%%%%%%%%%%%%%%%%%%%%

%%%%%%%%%%%%%%%%%%%%%%%%%%%%%%%%%%%%%%%%%%%%%%%%%%%%%%%%%%%%%%%%%%%%%%%%%%%%%%
\subsection{UV Spectra}

The first two UV spectra displayed in Fig.~\ref{fig_grism} show strong features also 
seen in the HST spectra of SN~1999em (Baron et~al.\ 2000).  Even after 
smoothing the spectra still contain significant noise, as evidenced by features in the 
optical regions which are not apparent in contemporaneous ground-based optical 
spectra \citep{Pastorello_etal_2006}.  Because of this and the $\sim30\AA$ uncertainty 
in the wavelength scale, we do not want to overinterpret line identifications 
or velocities.  We can identify a few strong lines, namely H$\beta$, H$\gamma$, 
and MgII 2798 multiplet, whose strong and broad absorption features are marked.
Overlapping absorption lines, 
particularly of Fe{\sc iii}, Fe{\sc ii} and Mg{\sc ii} remove much of the UV light 
(see e.g. \citealp{Lucy87}, \citealp{Br87}, \citealp{ML93}, \citealp{DH05, DH06}).   
A complete census of the various ions and the spectral regions where they affect the
emergent light is computed for two epochs bracketing the first grism observation 
(see discussion and the accompanying Fig.~5 below), generated by performing 
a formal solution to a converged
CMFGEN model, but accounting only for the bound-bound transitions 
of a single ion. This neglects potential non-linear effects and line overlap but offers an
instructive illustration of the cumulative effects of line blanketing, disentangled between
all contributing species.
Notice how little line emission appears above the UV continuum, and that the resulting ``emission'' features
in the UV coincide with spectral regions of {\it reduced} line blanketing.
The later UV spectra are of lesser quality but show the UV continuum becoming increasingly fainter.

The broad absorption around 2700\AA\ is consistent with the formation of Mg{\sc II}\,2798\AA\ 
in the SN ejecta, i.e. at and above the SN photosphere at that time, and we associate it with the
absorption part of the corresponding P-Cygni profile of this strong line (see \S~3.3).  
This suggests that the UV 
emission from SN~2005cs originates from the photospheric layers of the fast expanding ejecta 
rather than from interaction with the circumstellar medium (CSM).  
SNe 1979C, 1980K and 1993J all showed blue-shifted emission from MgII and 
had smoother UV spectra (see e. g. \citealp{Panagia_etal_1980}, \citealp{Fransson_1984}, 
\citealp{Immler_etal_2005}, \citealp{Jeffery_etal_2006}, \& \citealp{Jeffery_etal_1994}).  
SN~1987A, on the 
other hand, exhibited a P-Cygni profile for MgII and no evidence for strong 
CSM interaction at early times (\citealp{Kirshner_etal_1987}; \citealp{Pun_etal_1995}).
The addition of our spectra in Fig. 3 here to Fig. 3 in \citet{Jeffery_etal_1994} 
gives a spectroscopic juxtaposition of the same SNe compared photometrically 
in our Fig. 4.

%%%%%%%%%%%%%%%%%%%%%%%%%%%%%%%%%%%%%%%%%%%%%%%%%%%%%%%%%%%%%%%%%%%%%%%%%%%%%%

%%%%%%%%%%%%%%%%%%%%%%%%%%%%%%%%%%%%%%%%%%%%%%%%%%%%%%%%%%%%%%%%%%%%%%%%%%%%%%
\subsection{Comparison with non-LTE model atmospheres}

   We report on a preliminary quantitative analysis of the UVOT photometric evolution
and optical ground-based spectroscopy covering the first two weeks after discovery.  
Detailed results of this study will be shown in a forthcoming paper (Dessart et al. 2006, in prep.).

   We employ the non-LTE model atmosphere code CMFGEN \citep{HM98, DH05}
and follow the same approach as \citet{DH06}
for their analysis of SN~1999em. The version of the code used assumes a spherically 
symmetric (1D), steady-state, chemically homogeneous, homologously expanding ejecta 
with a density power-law of exponent $n$. 
A key ingredient of our approach is that we set the inner boundary luminosity of the 
model (at a Rosseland optical depth of $\sim$50-100) so that the emergent synthetic flux matches 
the observed flux.
To do this, we require the distance to SN~2005cs, which we take as 8\,Mpc,  
the rough average of various distance estimates to the host galaxy M51 or  sources within it.
These distance estimates used H{\sc II} regions \citep[9.6\,Mpc;][]{Sandage_Tammann_1974}, young stellar associations 
\citep[6.91\,Mpc;][]{Georgiev_etal_1990}, planetary nebulae 
\citep[8.4\,Mpc, revised to 7.6\,Mpc;][]{Fel97, Ciardullo_etal_2002}, 
surface brightness fluctuations \citep[7.7\,Mpc;][]{Tonry2001}, the SEAM 
\citep[6.0\,Mpc;][]{Baron_etal_1996}, and more recently the EPM/SCM using SN~2005cs itself \citep[7.3\,Mpc;][]{TV06}. 
We redden the synthetic spectral energy distribution computed by CMFGEN using the
law of \citet{Car88} and $E(B-V)$=0.04, which offers the best fit to the observations, 
and renormalize them to the observed flux at 6000\AA\ (the flux offset
between synthetic and observed spectra is small and $\sim$10-20\%).

   We assume an (homogeneous) ejecta composition that differs somewhat from that of 
Dessart \& Hillier (2006) used for SN1999em. Indeed, O{\sc ii} lines present in 
the June 30 spectrum at $\sim$4600\AA\ required a higher oxygen abundance \citep{Baron_2006}. We obtain
satisfactory fits at all times using H/He$ = 5$, C/He$=0.0004$, N/He$=0.0013$ and
O/He$=0.0016$, in agreement with the recent determination of 
the average surface composition of galactic B-supergiants \citep{Cro06}, together
with a solar metallicity for the other elements.

\begin{figure*} 
\epsscale{1.2}
\plottwo{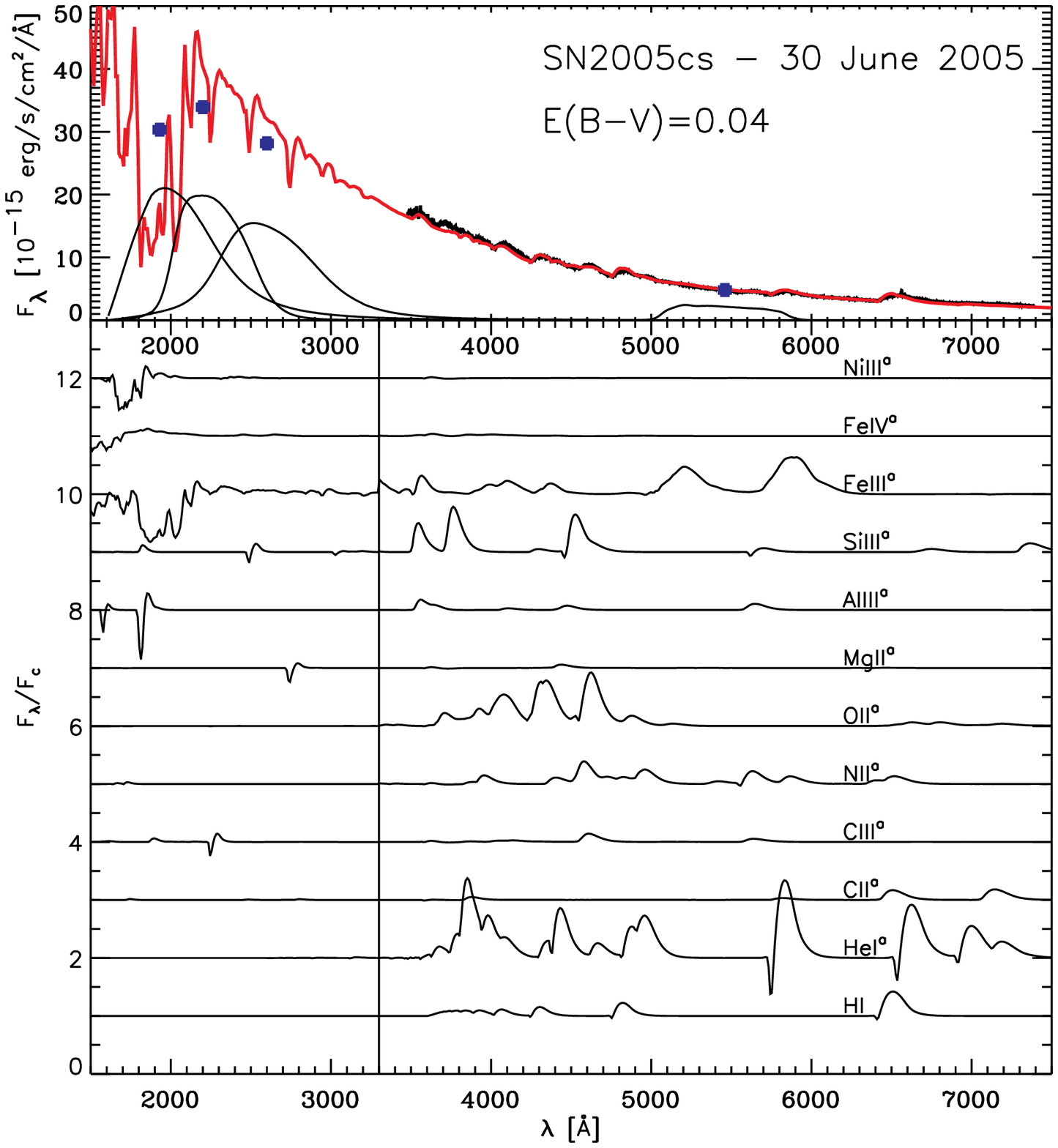}{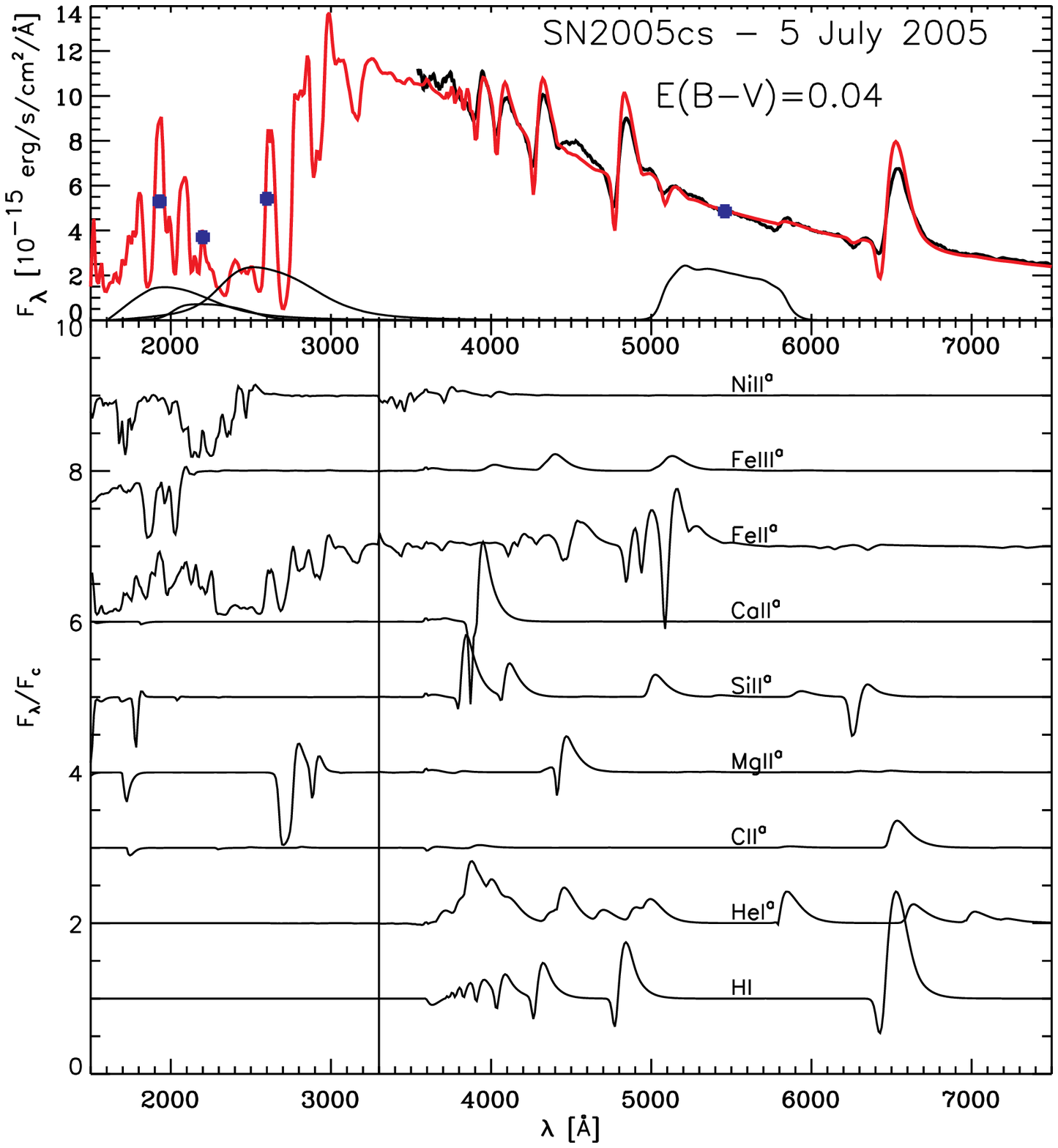}
\caption{
{\it Left}: Comparison between the photometric UVOT (blue crosses in color version), spectroscopic optical 
observations from CfA for SN~2005cs on the 30th of June 2006, and a reddened ($E(B-V) = 0.04$) 
synthetic spectrum computed with CMFGEN (red line in color version) using the procedure followed for the Type IIP 
SN~1999em and described in Dessart \& Hillier (2006). 
Model parameters are: $L_{\ast} =  2.7 \times 10^8 L_{\odot}$,
$T_{\rm phot} = 15750$K, $R_{\rm phot}= 2 \times 10^{14}$\,cm,
$v_{\rm phot}=$6900\,km\,s$^{-1}$, $\rho_{\rm phot} =  2.5 \times 10^{-13}$\,g\,cm$^{-3}$,
and $n=20$.
The synthetic flux, 20\% lower than observed for an adopted distance of 8\,Mpc, is re-normalized at 6000\AA.
UVOT fluxes are scaled so that the V-band flux matches the observed flux at 5460\AA.  
For both figures, we include in the lower panel the synthetic spectra obtained by
including bound-bound transitions only of the selected ions (and normalized by the continuum), 
thereby illustrating the sources
line blanketing, in particular in the UV range. Only the ions leaving a non-trivial mark are included.
Moreover, for ions marked with a superscript ''a'', we apply a magnification of 10 beyond
3300\AA\ to enhance the visibility of the weak features in the optical range.
{\it Right}: Same as left for the observations of SN~2005cs on the 5th of July 2005
and a reddened synthetic CMFGEN spectrum, whose corresponding model parameters are:
$L_{\ast} =  1.5 \times 10^8 L_{\odot}$,
$T_{\rm phot} = 8200$K, $R_{\rm phot}= 4.2 \times 10^{14}$\,cm,
$v_{\rm phot}=$5200\,km\,s$^{-1}$, $\rho_{\rm phot} =  6.4 \times 10^{-14}$\,g\,cm$^{-3}$,
and $n=10$.
The synthetic flux, 10\% lower than observed for an adopted distance of 8\,Mpc,
is re-normalized at 6000\AA, and the UVOT fluxes scaled to match the V band at 5460\AA. 
We also overplot the UVW2, UVM2, UVW1, and $V$ filter bandpasses, scaled in proportion 
to the observed flux and to avoid overlap with the spectra.
For both epochs, the (weak) UV brightness of the underlying H{\sc ii} at $\sim$80 days has been
subtracted. (See text for discussion.)
}
\label{fig_sed}
\end{figure*}

To document the early color evolution of the spectral energy distribution of SN~2005cs,
we focus on only two dates, June 30th and July 5th,
complementing UVOT observations with optical spectra obtained by the CfA SN Group (for June 30th) 
and Pastorello et~al.\ (2006; for July 5th) - the full time sequence will be shown in the follow-up analysis to
this paper. We show in Fig.~\ref{fig_sed} a comparison,
for these two dates, of the observed UVOT (blue crosses in color version) and optical spectra (black curve) of 
SN~2005cs with reddened synthetic spectra (red curve in color version) computed with CMFGEN.  At the 
bottom of each plot is a ladder plot of spectra (normalized by the continuum) showing the 
contribution of bound-bound transitions of individual ions.  
To fit the observations, we vary primarily the base radius/luminosity to modulate the
ionization state of the ejecta. This changes, in a non-linear way, the sources of
opacity thereby modifying line strengths.  It also changes the hardness of the spectral 
energy distribution (SED) 
at the thermalization layer (which lies above the inner boundary of the model), 
where it resembles that of a blackbody. Hence, this modifies the slope of the 
continuum as well. We also change the density exponent $n$, adopting a very high value
of 20 for the June 30th model.  This represents an 
extremely steep density decrease with radius, but is necessary in order to 
reproduce the very weak optical features.  A steep density profile was also required to fit 
early observations of SN~1993J \citep{Baron_etal_1995}.  A lower value of ten is used to reproduce the observations of July 5th.
We obtain satisfactory fits for a reddening of 0.04, which we adopt for all dates. Discordant
flux distributions between quasi-simultaneous observations of Pastorello et~al.\ (2006) and the CfA 
suggest that the flux is not very accurate below 4000\AA, likely due to the CfA spectra not being taken at the parallactic angle.  This is a region where spectra would best constrain 
the reddening and the ionization/temperature of the ejecta. The corresponding uncertainty in the slope
suggests that the reddening could be as low as 0, but could not be higher than 0.1 without 
setting stringent requirements on the spectral modeling. Either way, the reddening is low, with a value equal or 
lower than that adopted by Pastorello et~al.\ (2006), namely $E(B-V)=0.11$.  \citet{Baron_2006} 
also found the reddening to be small and adopted the galactic value 
for the extinction from the Schlegel dust maps ($E(B-V)=0.035$; \citealp{SFD1998}).

%%%%%%%%%%%%%%%%%%%%%%%%%%%%%%%%%%%%%%%%%%%%%%%%%%%%%%%%%%%%%%%%%%%%%%

%%%%%%%%%%%%%%%%%%%%%%%%%%%%%%%%%%%%%%%%%%%%%%%%%%%%%%%%%%%%%%%%%%%%%%

We find that the
density distribution flattens considerably in the continuum and line formation
region over that week, described here as a change in the density exponent from
$n=20$ to $n=10$.   This reproduces the changing optical line features, which
appear systematically weak  in the first spectrum but considerably stronger in
the second. Given the numerous constraints on our modeling, only varying the
density exponent can lead to the desired reduction of line fluxes in the first epoch.  The
corresponding flattening of the density profile over the course of one week
does not constitute a puzzle. As demonstrated by \citet{DH05}, line formation
in Type II SN is localized to a narrow spatial region above the photosphere
and, thus, no sizeable flux emission occurs beyond a few tens of percent beyond
the photospheric velocity. On the 5th of July 2005, $\sim$90\% of the
H$\alpha$ flux (the strongest optical line) received from SN~2005cs falls within
a range of line of sight velocities of $\pm$5000\kms, thus well below the
photospheric velocity on the 30th of June for which we adopt the steeper
density slope. Another point (Woosley, priv. comm.) is that the ejecta swept up
by the shock  remains in a dynamical phase up to one week  after breakout.
Homologous expansion is not reached until after about one week.  This suggests
that interpreting and comparing ejecta velocities inferred spectroscopically 
during the first week of explosion is not straightforward.

Beside a flattening of the density distribution, we also find that the ionization/temperature changes 
significantly between these two dates, following cooling through expansion
of the ejecta, as well as radiation from the surface layers. We infer a drop of the photospheric 
temperature from 15750\,K down to 8200\,K in that period. 
(We define the photosphere as the ejecta location where the inward integrated optical depth at 
5000\AA, including bound-free, free-free, and electron-scattering opacity processes, is equal 
to two-thirds.) The corresponding constraints from the observations are the fast changing UV 
flux/magnitude and the strengthening of line blanketing in the UV and, more modestly, in the optical.
Line identifications apply here the same way as for SN~1999em (Dessart \& Hillier 2006), 
but with a few alterations.
The first spectrum is likely closer to the explosion date (see Dessart et~al.\ 2006), showing 
much weaker lines and a steeper spectrum than for SN~1999em. The ejecta of SN~2005cs is also 
slower, so that we not only confirm the presence of N{\sc ii} lines (Dessart \& Hillier 2006) 
but also of O{\sc ii} at 4600\AA.
The Si{\sc ii} 6355\AA\ line is also well resolved, while it overlapped with the H$\alpha$ trough in SN1999em 
(Dessart \& Hillier 2006; Leonard et~al.\ 2002b) or SN1999gi (Leonard et~al.\ 2002a).

\begin{figure*}
\epsscale{1.3}
\plottwo{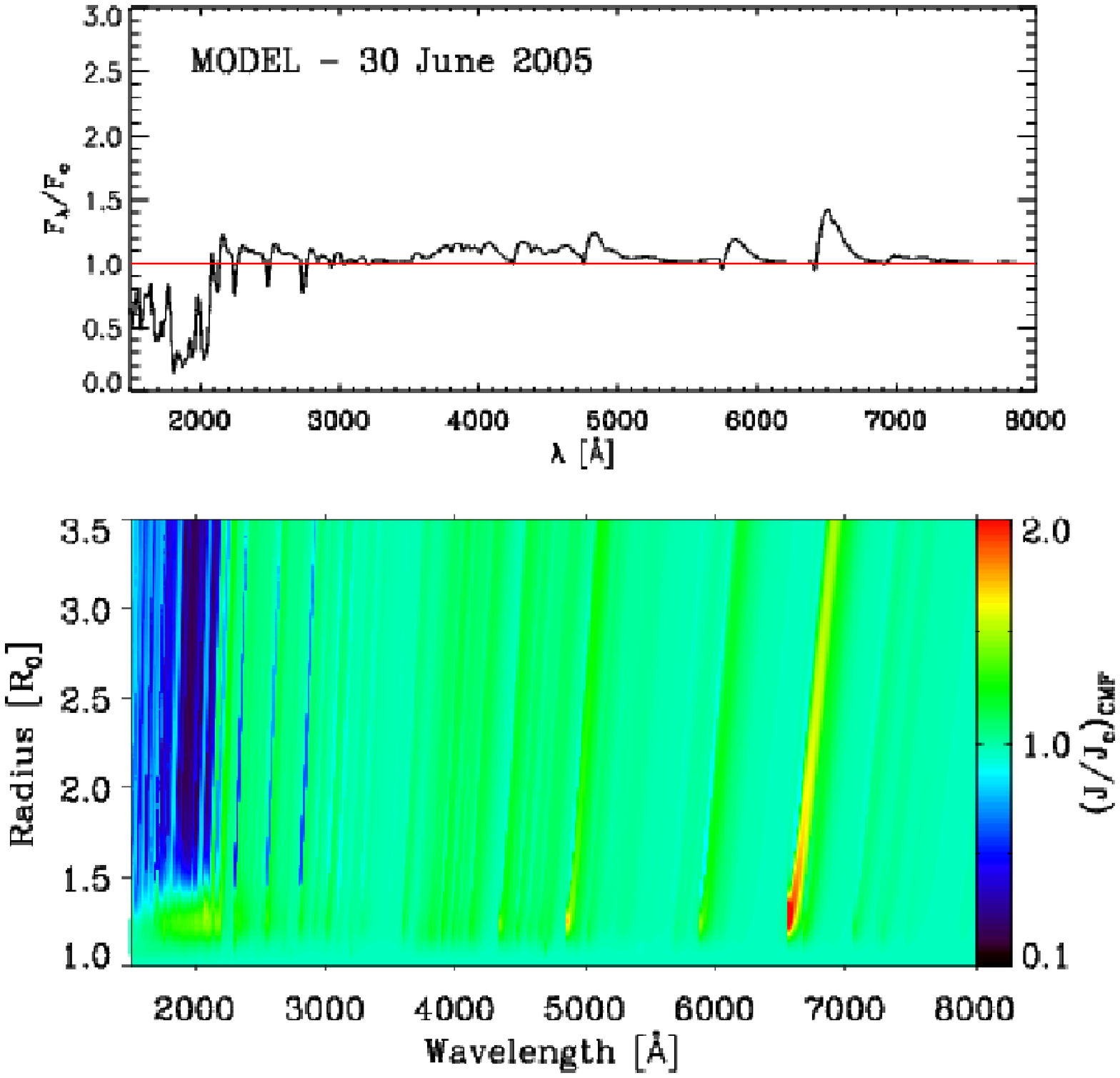}{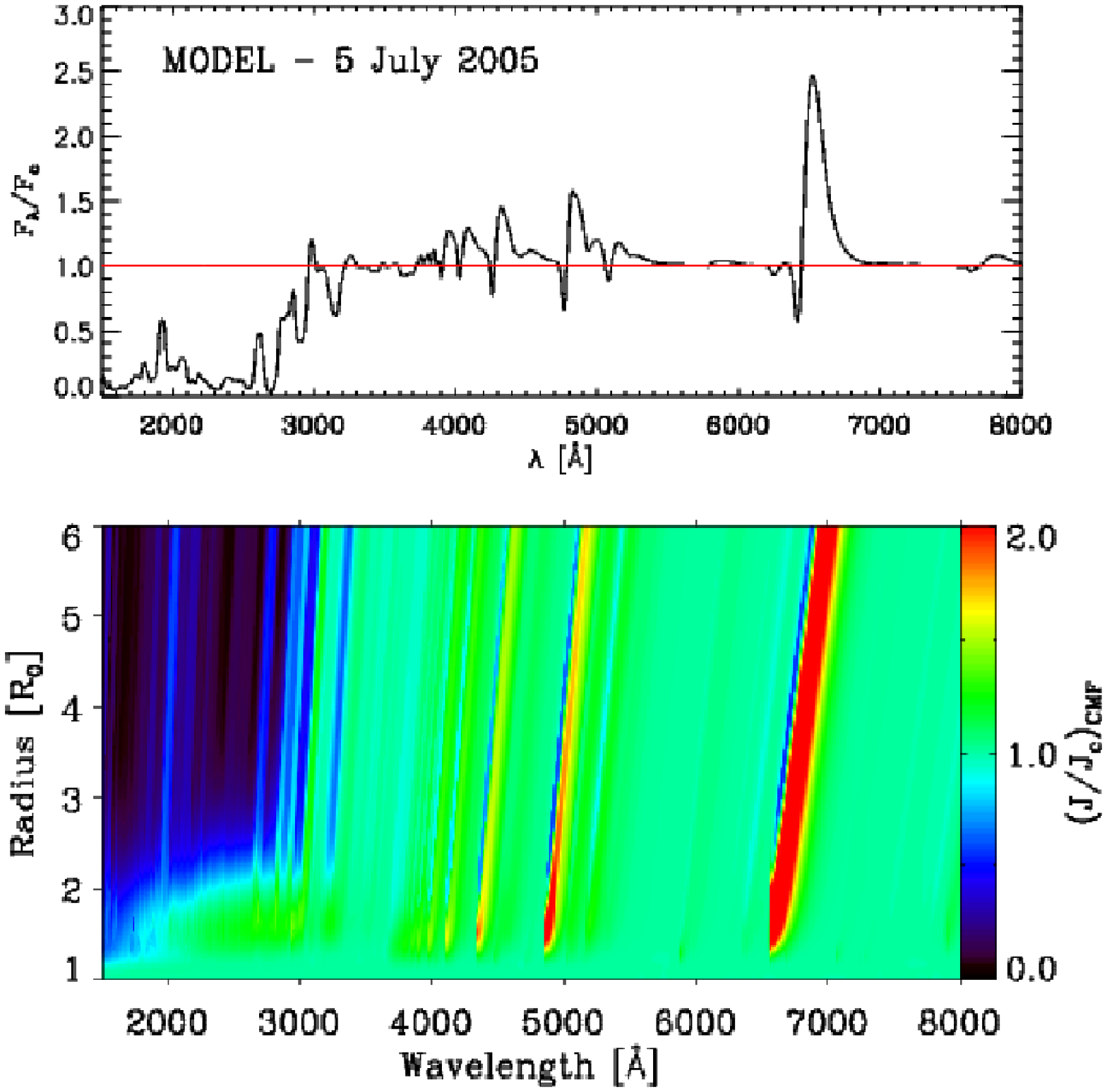}
\caption{{\it Left}: Radial variation (in the comoving frame) of the mean intensity normalized to 
the continuum mean intensity over the UV and optical ranges (bottom panels) for the June 30th CMFGEN model.
In the top panel, we show the synthetic emergent spectrum, normalized to the continuum spectrum.
% show the same ratio but in the observer's frame and at the maximum CMFGEN grid radius. 
{\it Right}: Same as left, but this time for the July 5th CMFGEN model.}
\label{fig_lb}
\end{figure*}

The temporal evolution of the flux distribution reflects first the cooling of the photosphere.
At the base of our two CMFGEN models (corresponding to a Rosseland optical depth of 70 and 
73, or a radius of 1.6 and  4.176 $\times$ 10$^{14}$\,cm for the two selected dates), 
the electron temperature
varies from 40750\,K down to 23670\,K, and the corresponding flux distributions, which match 
blackbodies at such depths, peak at 710\AA\ and 1220\AA. We thus see that the intrinsic photon distribution
softens significantly, although it still peaks in the UV.
The flux distribution that emerges peaks further to the red due to the intervening blocking
of light in the UV, between the thermalization layer where the photon distribution is characterized
to layers where it escapes freely. We illustrate this effect in Fig.~\ref{fig_lb}
by showing the radial variation in the comoving frame (the radiation field appears redshifted with increasing radius) 
of the mean intensity ($F_\lambda$) normalized to the continuum mean intensity ($F_{\rm c}$; bottom panels),
for the June 30th (left) and the July 5th (right) models. Here the continuum mean intensity is 
computed by solving the radiative transfer equation for the processes associated with the continuum only.  
At the base $F_\lambda/F_{\rm c}$ is unity, but as we cross the
photosphere (at $\sim$1.4$R_{\rm 0}$ for June 30 and $\sim$2$R_{\rm 0}$ for July 5, where $R_{\rm 0}$ is the base model radius), this ratio suddenly drops below
unity in the UV, while showing the appearance of many lines in the optical.  
At such early times, optical lines are mostly of H{\sc i} and He{\sc i}, with weaker contributions from 
N{\sc ii} and O{\sc ii}.
In the top panel, we show the emergent synthetic spectrum (normalized to the continuum) for comparison.
Hence, we see that line blanketing plays a strong role in the UV, but it operates on a photon distribution 
that becomes more and more depleted in the UV as time progresses. Our
model atmosphere calculations support a purely photospheric origin for 
the observed flux, both for the UV and the optical ranges, and thus do 
not suggest any sizeable CSM contribution (ignoring the underlying HII 
region contribution).
These two combined effects, the reddening of the continuum and the strengthening of the 
line blanketing, explain the color evolution of SN~2005cs and in particular confirm the
``thermal'', photospheric origin of this radiation. Work is underway to interpret quantitatively
the absolute flux levels observed, with links to the explosion energy and the ejecta kinematics. 
%%%%%%%%%%%%%%%%%%%%%%%%%%%%%%%%%%%%%%%%%%%%%%%%%%%%%%%%%%%%%%%%%%%%%%%%%%%%%%

%%%%%%%%%%%%%%%%%%%%%%%%

\subsection{X-Ray Results}
\label{results_xray}

The upper limit from the merged 26.8~ks XRT observation (after screening out the 
day of the X-Ray flash) corresponds to
an unabsorbed X-ray flux and luminosity of
$f_{0.2-10}\ls 9.0 \times 10^{-14}~{\rm ergs~cm}^{-2}~{\rm s}^{-1}$ and
$L_{0.2-10}\ls 8.4 \times 10^{38}~{\rm ergs~s}^{-1}$, respectively for
an assumed thermal plasma spectrum with a temperature of 10~keV, an
absorbing foreground column density of
$1.6 \times 10^{20}~{\rm cm}^{-2}$ \citep{DL} and a
distance of 8~Mpc.  SNe IIP have been detected in X-rays in the past (SN
1999em, \citealp{Pooley_etal_2002}; SN~1999gi, \citealp{Schlegel_2001}; SN2004dj, 
\citealp{Pooley_Lewin_2004})
but usually at lower luminosities (typically around $10^{38}~{\rm ergs~s}^{-1}$) than
our upper limits for SN~2005cs.  But these observations do rule out an X-ray bright SN 
more luminous than SN~1994W ($8\times 10^{38}~{\rm ergs~s}^{-1}$), a IIP with optical evidence for CSM 
interaction \citep{Schlegel_1999}.  The typically faint X-ray luminosity is
attributed to a low density CSM surrounding the
progenitor stars \citep{Schlegel_2001}.  The presence of X-rays would likely 
indicate interaction with the CSM which would also affect the UV luminosity.  
Thus the lack of X-rays is consistent with the photospheric origin of 
the UV radiation discussed previously.

We calculated the upper limit to the mass-loss rate of the progenitor
assuming a shock velocity of 10,000~${\rm km~s}^{-1}$ 
and following the description by Immler, Wilson \& Terashima (2002).
A 3-sigma upper limit of
$\dot{M}\ls 1 \times 10^{-5}~M_{\odot}~{\rm yr}^{-1}~(v_{\rm w}/10~{\rm km~s}^{-1})$, 
where $v_{\rm w}$ corresponds to the wind velocity, 
is obtained for a median date of the XRT observations around
$12\pm1$ days after the outburst.  The upper limit to the mass loss-rate is 
consistent with those of core-collapse SN progenitors, which are in the range 
$10^{-6}$ to $10^{-4} M_\odot~{\rm yr}^{-1}$ \citep{Immler_Lewin_2003} and 
$\sim10$ times smaller than that of highly interacting SNe such as 
SN~1979C \citep{Immler_etal_2005}.

%%%%%%%%%%%%%%%%%%%%%%%%%%%%%%%%%%%%%%%%%%%%%%%%%%%%%%%%%%%%%%%%%%%%%%%%%%%%%%
\section{Summary\label{SECTION:conc}}

SN~2005cs is the first SN IIP with a well observed UV light curve, and is 
an important addition to the SNe II observed in the UV by IUE and HST.
The rapid drop in the UV shows the importance of quick response observations, 
as the emergent flux is dominated by the UV for only about a week after the 
explosion.  The agreement of our CMFGEN models with the 
UVOT photometry and optical spectra (for a unique set of SN ejecta parameters), 
the identification of absorption lines in the 
UV (e.g., Mg{\sc ii}\,2798\AA) and in the optical (e.g., H{\sc i} Balmer lines with
comparable, Doppler-broadened widths), and the non-detection of X-rays 
all point to a photospheric origin of the UV emission 
and a lack of strong CSM interaction.  The early UV photometry presented 
in this paper demonstrates the 
temporal effects of photospheric cooling and line blanketing which can be 
reproduced by the non-LTE model atmosphere code CMFGEN.   The effort to understand these effects 
will be continued by increasing the sample of UV light curves and spectra of 
SNe II with {\sl Swift} in order to 
provide constraints for the spectroscopic modeling of multiple objects.
These observations will also provide X-ray observations probing earlier times 
than have been studied in the past.

%%%%%%%%%%%%%%%%%%%%%%%%%%%%%%%%%%%%%%%%%%%%%%%%%%%%%%%%%%%%%%%%%%%%%%%%%%%%%%%%%

\acknowledgements

This work made use of the NASA/IPAC Extragalactic Database.  This work is 
supported at Penn State by NASA contract NAS5-00136.  Supernova studies at Harvard 
University are supported by NSF grant AST06-06772.  RPK acknowledges support from the Kavli Institute for Theoretical Physics through NSF grant PHY99-07949.  The F. L. Whipple Observatory is operated by the Smithsonian Astrophysical Observatory.

%%%%%%%%%%%%%%%%%%%%%%%%%%%%%%%%%%%%%%%%%%%%%%%%%%%%%%%%%%%%%%%%%%%%%%%%%%%%%%%%%

%%%%%%%%%%%%%%%%%%%%%%%%%%%%%%%%%%%%%%%%%%%%%%%%%%%%%%%%%%%%%%%%%%%%%%%%%%%%%%%%%

\onecolumn

%%%%%%%%%%%%%%%%%%%%%%%%%%%%%%%%%%%%%%%%%%%%%%%%%%%%%%%%%%%%%%%%%%%%%%%%%%%%%%
%% The values (usually only l,r and c) in the last part of
%% \begin{deluxetable}{} command tell LaTeX how many columns
%% there are and how to align them.
\begin{deluxetable}{ccccc}
%\tablewidth{8cm}
%\tabletypesize{\scriptsize}
\tablecaption{{\sl Swift} UVOT Filter Characteristics}
%\tablenum{1}
\tablehead{
\colhead{Filter} & \colhead{$\lambda_{\rm c}$} & \colhead{FWHM} & \colhead{Zeropoint}  & \colhead{Flux Density} \\ 
\colhead{}       & \colhead{(\AA)} & \colhead{(\AA)} & \colhead{(mag)}  & \colhead{(10$^{-16}$~ergs cm$^{-2}{\rm~counts^{-1}}$)} 
}
%% All data must appear between the \startdata and \enddata commands
\startdata
UVW2 & 1880 & 760 & $17.77 \pm 0.20$ & $6.04 \pm 0.42$\\
UVM2 & 2170 & 510 & $17.29 \pm 0.23$  & $6.89 \pm 0.96$\\
UVW1 & 2510 & 700 & $17.69 \pm 0.20$  & $3.52 \pm 0.07$\\
$U$ & 3450 & 875 & $18.38 \pm 0.23$  & $1.49 \pm 0.06$\\
$B$ & 4390 & 980 & $19.16 \pm 0.12$  & $1.31 \pm 0.14$\\
$V$ & 5440 & 750 & $17.88 \pm 0.09$  & $2.24 \pm 0.12$\\
\enddata
%% Include any \tablenotetext{key}{text}, \tablerefs{ref list},
%% or \tablecomments{text} between the \enddata and 
%% \end{deluxetable} commands
%% No \tablecomments indicated
\tablecomments {$\lambda_{\rm c}$ refers to the central wavelength of the filter's 
effective area curve.  The flux density is a multiplicative conversion from count rate to flux density 
assuming a Vega spectrum.  The zeropoints and flux 
density factors correspond to a source with a Vega-like spectrum and a count rate of one count
per second.  }
%% No \tablerefs indicated
\end{deluxetable}

%\clearpage

%%%%%%%%%%%%%%%%%%%%%%%%%%%%%%%%%%%%%%%%%%%%%%%%%%%%%%%%%%%%%%
%\clearpage

%% The values (usually only l,r and c) in the last part of
%% \begin{deluxetable}{} command tell LaTeX how many columns
%% there are and how to align them.
\begin{deluxetable}{lcccccccccccc}

%% Rotate to a landscape orientation
%\rotate

%% Over-ride the default font size
%% Use Default (12pt)

%% Use \tablewidth{?pt} to over-ride the default table width.
%% If you are unhappy with the default look at the end of the
%% *.log file to see what the default was set at before adjusting
%% this value.

%% This is the title of the table.
\tablecaption{{\sl Swift} UVOT Photometry of SN~2005cs}

%% This command over-rides LaTeX's natural table count
%% and replaces it with this number.  LaTeX will increment
%% all other tables after this table based on this number
%\tablenum{2}

%% The \tablehead gives provides the column headers.  It
%% is currently set up so that the column labels are on the
%% top line and the units surrounded by ()s are in the
%% bottom line.  You may add more header information by writing
%% another line between these lines. For each column that requries
%% extra information be sure to include a \colhead{text} command
%% and remember to end any extra lines with \\ and include the
%% correct number of &s.
\tablehead{ \colhead{JD} & \colhead{UVW2}  & \colhead{UVM2}  & \colhead{UVW1} &  \colhead{$V$}  \\
}

%% All data must appear between the \startdata and \enddata commands
\startdata

2453552.4 & $ 13.46 \pm 0.06 $ & $ 13.01 \pm 0.10 $ & $ 12.89 \pm 0.02 $ & $ 14.55 \pm 0.04 $ \\
2453555.1 & $	14.43 \pm 0.06 $ &  \nodata           &  \nodata           & $ 14.47 \pm 0.10 $ \\
2453557.3 & $	15.02 \pm 0.06 $ & $ 14.97 \pm 0.10 $ &  \nodata           & $ 14.53 \pm 0.04 $ \\
2453557.9 & $	15.22 \pm 0.06 $ & $ 15.20 \pm 0.10 $ & $ 14.52	\pm 0.03 $ & $ 14.55 \pm 0.06 $ \\
2453558.8 & $	15.75 \pm 0.07 $ & $ 15.62 \pm 0.11 $ & $ 14.80	\pm 0.03 $ & $ 14.65 \pm 0.04 $ \\
2453560.0 & $	16.02 \pm 0.07 $ & $ 16.04 \pm 0.11 $ & $ 15.23	\pm 0.03 $ & $ 14.57 \pm 0.04 $ \\
2453560.6 & $	16.26 \pm 0.07 $ & $ 16.14 \pm 0.11 $ & $ 15.43	\pm 0.04 $ & $ 14.58 \pm 0.06 $ \\
2453562.0 & $	16.33 \pm 0.07 $ & $ 16.26 \pm 0.11 $ & $ 15.72	\pm 0.04 $ & $ 14.55 \pm 0.06 $ \\
2453562.8 & $	16.62 \pm 0.08 $ & $ 16.22 \pm 0.11 $ & $ 15.78	\pm 0.05 $ & $ 14.63 \pm 0.06 $ \\
2453565.0 & $	16.46 \pm 0.08 $ & $ 16.36 \pm 0.12 $ & $ 15.94	\pm 0.06 $ & $ 14.55 \pm 0.08 $ \\
2453567.0 & $	16.50 \pm 0.07 $ & $ 16.24 \pm 0.11 $ & $ 15.94	\pm 0.04 $ & $ 14.65 \pm 0.05 $ \\
2453571.0 & $	16.62 \pm 0.08 $ & $ 16.49 \pm 0.12 $ & $ 16.11	\pm 0.05 $ & $ 14.73 \pm 0.08 $ \\
2453573.5 & $	16.50 \pm 0.08 $ & $ 16.32 \pm 0.12 $ & $ 16.23	\pm 0.06 $ & $ 14.76 \pm 0.09 $ \\
2453628.5 & $	16.67 \pm 0.07 $ & $ 16.39 \pm 0.11 $ & $ 16.44	\pm 0.04 $ & $ 14.86 \pm 0.03 $ \\
\enddata

%% Include any \tablenotetext{key}{text}, \tablerefs{ref list},
%% or \tablecomments{text} between the \enddata and
%% \end{deluxetable} commands

%% General table comment marker
\tablecomments{The observation time is given as the Julian Date (JD) of the middle of the 
set of exposures.  
These values have not been corrected for extinction.  The errors given are $1\sigma$ statistical errors.}

%% No \tablerefs indicated

\end{deluxetable}

%%%%%%%%%%%%%%%%%%%%%%%%%%%%%%%%%%%%%%%%%%%%%%%%%%%
%% The values (usually only l,r and c) in the last part of
%% \begin{deluxetable}{} command tell LaTeX how many columns
%% there are and how to align them.
%\clearpage
\begin{deluxetable}{cccccccc}

%% Keep a portrait orientation

%% Over-ride the default font size
%% Use Default (12pt)

%% Use \tablewidth{?pt} to over-ride the default table width.
%% If you are unhappy with the default look at the end of the
%% *.log file to see what the default was set at before adjusting
%% this value.

%% This is the title of the table.
\tablecaption{{\sl Swift} UVOT Grism Observations}

%% This command over-rides LaTeX's natural table count
%% and replaces it with this number.  LaTeX will increment 
%% all other tables after this table based on this number
%\tablenum{1}

%% The \tablehead gives provides the column headers.  It
%% is currently set up so that the column labels are on the
%% top line and the units surrounded by ()s are in the 
%% bottom line.  You may add more header information by writing
%% another line between these lines. For each column that requries
%% extra information be sure to include a \colhead{text} command
%% and remember to end any extra lines with \\ and include the 
%% correct number of &s.
\tablehead{\colhead{Observation Sequence} & \colhead{Date (UT)} &  \colhead{Exposure} & \colhead{Grism} \\ 
\colhead{} & \colhead{}  & \colhead{(Seconds)} & \colhead{} } 

%% All data must appear between the \startdata and \enddata commands
\startdata
00030083007 & 3 Jul 2005 &  2078.1 & V \\
00030083008 & 3 Jul 2005 &  2078.5 & UV \\
00030083012 & 6 Jul 2005 &  2137.9 & UV \\
00030083017 & 8 Jul 2005 &  2017.7 & UV \\
00030083023 & 11 Jul 2005 &  2077.7 & UV \\
00030083027 & 13 Jul 2005 &  1788.3 & UV \\
00030083035 & 19 Jul 2005 &  1955.2 & UV \\
\enddata

%% Include any \tablenotetext{key}{text}, \tablerefs{ref list},
%% or \tablecomments{text} between the \enddata and 
%% \end{deluxetable} commands

%% No \tablecomments indicated

%% No \tablerefs indicated

\end{deluxetable}
%%%%%%%%%%%%%%%%%%%%%%%%%%%%%%%%%%%%%%%%%%%%%%%%%%%%%%%%%%%%%%%%%%%%%%%%%%%%%%

\clearpage
%\clearpage

%\clearpage
%\clearpage
%%%%%%%%%%%%%%%%%%%%%%%%%%%%%%%%%%%%%%%%%%%%%%%%%%%%%%%%%%%%%%%%%%%%%%%%%%%%%%

%%%%%%%%%%%%*******************************

%\clearpage
%\clearpage

%%%%%%%%%%%%%%%%%%%%%%%%%%%%%%%%%%%%%%%%
%%%%%%%%%%%%%%%%%%%%%%%%%%%%%%%%%%%%%%%%%%%%%%%%%%%%%%%%%%%%%%%%%%%%%%%%%%%%%%
%\clearpage
%\clearpage

%%%%%%%%%%%%%%%%%%%%%%%%%%%%%%%%%%%%%%%%%%%%%%%%%%%%%%%%%%%%%%%%%%%%%%%%%%%%%%
%\clearpage
%%%%%%%%%%%%%%%%%%%%%%%%%%%%%%%%%%%%%%%%%%%%%%%%%%%%%%%%%%%%%%%%%%%%%%

%\clearpage
%%%%%%%%%%%%%%%%%%%%%%%%%%%%%%%%%%%%%%%%%%%%%%%%%%%%%%%%%%%%%%%%%%%%%%

\end{document}